\documentclass[doublecol]{epl2}

\title{Non-affine response: jammed packings versus spring networks}

\author{Wouter G. Ellenbroek\inst{1,2} \and Zorana
Zeravcic\inst{2} \and Wim van Saarloos\inst{2}  \and Martin van
Hecke\inst{3}} \shortauthor{Wouter Ellenbroek \etal}

\institute{\inst{1} Department of Physics and Astronomy,
University of Pennsylvania, Philadelphia, PA 19104-6396, USA\\
\inst{2} Instituut-Lorentz,  Universiteit Leiden, Postbus
9506, 2300 RA Leiden, The Netherlands\\
\inst{3} Kamerlingh Onnes Lab, Universiteit Leiden, Postbus
9504, 2300 RA Leiden, The Netherlands}


\providecommand{\vc}[1]{\mathbf{#1}}
\newcommand{\uij}{\vc{u}_{ij}}
\newcommand{\rij}{\vc{r}_{ij}}
\newcommand{\uparl}{u_\parallel}
\newcommand{\uperp}{u_\perp}
\newcommand{\dm}{\mathcal{M}}
\usepackage{graphicx}
\usepackage[]{amsmath,amssymb,amsfonts}
\usepackage{color}
\usepackage[normalem]{ulem}

\abstract{We compare the elastic response of spring networks whose contact
geometry is derived from real packings of frictionless discs, to networks
obtained by randomly cutting bonds in a highly connected network derived from a
well-compressed packing. We find that the shear response of packing-derived
networks, and both the shear and compression response of randomly cut networks,
are all similar: the elastic moduli vanish linearly near jamming, and
distributions characterizing the local geometry of the response scale with
distance to jamming. Compression of packing-derived networks is exceptional:
the elastic modulus remains constant and the geometrical distributions do not
exhibit simple scaling. We conclude that the compression response of jammed
packings is anomalous, rather than the shear response.}

\begin{document}

\maketitle
\date{\today}

The jamming transition governs the onset of rigidity in disordered
media as diverse as foams, colloidal suspensions, granular media
and glasses~\cite{jamnature}. While jamming in general is
controlled by a combination of density, shear stress and
temperature, most progress has been made for frictionless soft
spheres that interact through purely repulsive contact forces, and
that are at zero temperature and zero load~\cite{epitome,silbertPRL05,silbertPRE06,wyartEPL,wyartPRE,respprl}.
This simple model applies to static foams or emulsions~\cite{bolton,durianbubble}, and represents a simplified version
of granular media, if one ignores friction~\cite{maksePRL,ellakjam} and nontrivial grain shapes~\cite{donev04,wouterse07,bulbul09,zorana09}.

From a theoretical point of view, this model is ideal for two
reasons. First, it exhibits a well defined jamming point, \emph{``point J''},
which in the limit of large system sizes, occurs at
a well-defined density $\phi=\phi_c$~\cite{epitome}.
Here the system is a disordered packing of frictionless undeformed
spheres, which is marginally stable and isostatic, i.e., its
contact number (average number of contacts per particle) $z$
equals $z_\mathrm{iso}=2d$ in $d$ dimensions~\cite{epitome,moukarzelPRL98}. Second, in recent years it has been
uncovered that the mechanical and geometric properties of such
jammed packings exhibit a number of non-trivial power law scalings
as a function of the distance to the jamming point:
\emph{(1)} The excess contact number $\Delta z:= z-z_\mathrm{iso}$
scales as $(\phi- \phi_c)^{1/2}$~\cite{durianbubble,maksePRL,epitome,wyartPRE}; \emph{(2)} The ratio
of shear ($G$) and bulk ($K$) elastic moduli vanishes at point $J$
as $G/K \sim \Delta z$~\cite{epitome}.

\begin{figure}[t]
\includegraphics[width=8.4cm]{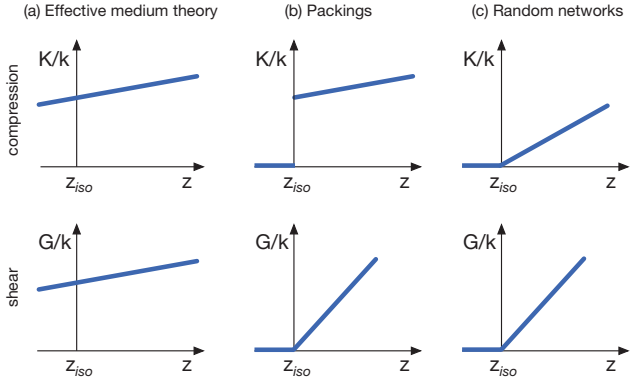}
\caption{Schematic comparison of the variation of shear ($G$) and bulk
($K$) elastic moduli as function of distance to jamming. (a) In
effective medium theory, all elastic moduli are simply of the
order of the local spring constant $k$, and moreover, the theory
does not account for whether the packing is rigid or not. (b) In
jammed packings of harmonic particles, the bulk modulus $K$
remains constant down to the jamming transition, where it vanishes
discontinuously, whereas the shear modulus $G$ vanishes linearly
in $\Delta z$. (c) In random networks of elastic springs, both
elastic moduli vanish linearly with $\Delta z$. }
\label{fig:3cases}
\end{figure}

The latter behavior --- a shear rigidity which becomes much
smaller than the compression modulus as the jamming point is
approached -- is in many ways surprising. It also differs markedly
from what is found in two simplified models of jammed systems,
effective medium theory (EMT) and random elastic networks, as is
illustrated schematically in fig.~\ref{fig:3cases} for the simple
case of harmonic particles. EMT predicts that the elastic moduli
vary smoothly through the isostatic point where $\Delta z=0$ and
that the moduli are of order of the local spring constant $k$.
This is because effective medium theory is essentially ``blind''
to local packing considerations and isostaticity.
Thus, besides failing to capture the vanishing of $G$ near jamming,
its prediction for the bulk modulus fails spectacularly as well:
it predicts finite rigidity below isostaticity.

The failure of EMT to describe elasticity near jamming motivated
earlier suggestions that elasticity of jammed packings might be
captured by random networks of springs
--- this problem is known as rigidity percolation~\cite{thorpe95,thorpe96,bolton,mason95emulsions}. However, in such
random spring networks, \emph{both} $G$ \emph{and} $K$ are expected
to go to zero as $k \Delta z$, as fig.~\ref{fig:3cases}c
illustrates~\cite{thorpe95}.

Thus, while from the point of view of effective medium theory the
\emph{shear rigidity} of jammed packings behaves anomalously,  from
the point of view of rigidity percolation, the \emph{compression
modulus} behaves unexpectedly.
What sets jammed packings apart from either of these two limiting
models? How to understand the difference in terms of the local
packing or response? Is the difference with rigidity percolation
visible in the scaling behavior of the response of packings? These
are issues we aim to clarify in this paper. Our approach will
hinge on characterizing the elastic response at the level of the
bonds. After all, the elastic moduli characterize changes in
elastic energy $\Delta E$ under deformations, and $\Delta E$
simply is a sum of the changes in elastic energy of all contacts
(bonds) in the system.

By probing the nature of the local response of packing-derived and
randomly cut networks, we find that we can distinguish two cases.
In the ``generic'' case, all geometrical characterizations exhibit
simple scaling and the elastic moduli scale as $\Delta z$ --- this
describes shear and bulk deformations of randomly cut networks, as
well as shear deformations  of packing-derived networks. Packing
derived networks under compression form the ``exceptional'' case:
the fact that the compression modulus remains of order $k$ near
jamming is reflected in the fact that various characteristics of
the local displacements do \emph{not} exhibit pure scaling. We
connect these findings to recent theoretical work by
Wyart~\cite{wyartthesis,wyart08moduli}.

\begin{figure}[t]
\includegraphics[width=8.4cm]{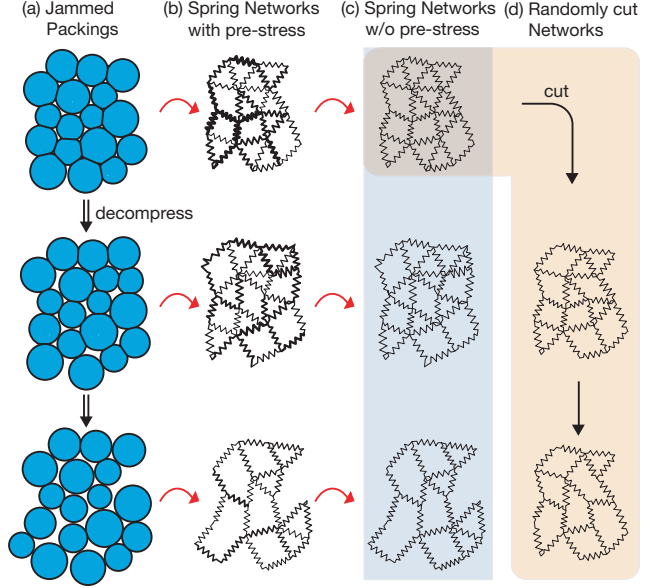}
\caption{Two families of spring networks --- see text for details.}
\label{fig:p2n}
\end{figure}

\section{Linear Response}
All numerical results presented in this paper concern
{quasistatic} linear response of systems to global shear or
compressional forcing. First we generate, for a range of
pressures, ensembles of 50 two-dimensional jammed packings
of 1024 frictionless particles with one-sided harmonic forces ($k=1$)
using a Molecular Dynamics simulation (for details, see~\cite{ellakwave}). Our linear response calculations are based on
the dynamical matrix. We decompose, for linear deformations, the
relative displacement $\uij$ of neighboring particles $i$ and $j$
in components parallel $(\uparl)$ and perpendicular $(\uperp)$ to
$\rij$, where $\rij$ connects the centers of particles $i$ and
$j$. In these terms the change in energy takes a simple form~\cite{alexander,wyartEPL},
\begin{equation}
\label{dynmatdef} \Delta E=\frac12 \sum_{i,j} \frac{k}{2}
\left(~ u^2_{\parallel,ij} -\frac{f_{ij}}{k ~r_{ij}}u_{\perp,ij}^2\right)
 ~.
\end{equation}
The dynamical matrix $\dm_{ij,\alpha\beta}$ is obtained by rewriting eq.~(\ref{dynmatdef}) in terms
of the independent variables, $u_{i,\alpha}$, as
\begin{equation}
\label{dynmatdef2} \Delta E=\frac12
\dm_{ij,\alpha\beta}~u_{i,\alpha}~u_{j,\beta}
~.
\end{equation}
Here $f_{ij}$ are contact forces, $k$ denotes the stiffness of the
springs\footnote{For our harmonic potential, $k=1$ for each contact,
but the procedure works equally well for more general potentials,
for which $k_{ij}$ is simply the value of the second derivative of the
potential, evaluated at the initial distance $r_{ij}$.},
$\dm$ is a $dN\times dN$ matrix with $N$ the number of
particles, 
indices $\alpha,\beta$ label the coordinate axes,
and the summation convention is used. The dynamical matrix contains
all information on the elastic properties of the system, and in
particular describes the linear response to external forcing
$f^\mathrm{ext}_{i,\alpha}$ as~\cite{leonforte,respprl}:
\begin{equation}
\dm_{ij,\alpha\beta}~u_{j,\beta}=f^\mathrm{ext}_{i,\alpha}~.\label{responseeq}
\end{equation}

\section{Two Families of Spring Networks}
We
start by noting that the analysis of the linear response of jammed
packings of particles with one-sided harmonic interactions
(fig.~\ref{fig:p2n}a) is exactly equivalent to that of networks of
appropriately loaded harmonic springs (fig.~\ref{fig:p2n}b), with
the nodes of the network given by the particle centers and the
geometry and forces of the spring network determined by the force
network of the packing.

In all that follows, we ignore the pre-stress term $
\frac{f_{ij}}{k ~r_{ij}}u_{\perp,ij}^2$ which is subdominant near
jamming ---
we have checked that its exclusion does not affect the results~\cite{wyartPRE}. The system without pre-stress is equivalent to a
``neutral'' spring network where all
{contacts are replaced by springs at their equilibrium length so}
that $f_{ij}=0$ {for all contacts} (fig.~\ref{fig:p2n}c). For such a neutral
spring network the dynamical matrix becomes particularly simple,
as its only non-zero elements are simply given by geometry and by
the bond strengths $k$ of each bond.

We follow two routes to approach the (un)jamming transition by
lowering the contact number in the neutral networks. In the first
route, we map jammed packings under increasingly low pressure
(fig.~\ref{fig:p2n}a) to neutral, packing-derived spring networks
(fig.~\ref{fig:p2n}c)
--- the geometry of these networks and the contacts that are
broken when point J is approached are thus set by real packings {that
were created using the MD protocol described in Ref.~\cite{ellakwave}}.
In the second route, we start from {a neutral spring network that is
obtained from} a heavily compressed jammed
packing for which $z \approx 5.98$. We then create randomly cut
networks with lower contact number by randomly removing springs~\cite{wyart08}, making sure that we do not create local
disconnected patches and that each node in the network remains
connected by at least three springs (fig.~\ref{fig:p2n}d)
--- the geometry of these networks becomes increasingly random. {Note that no relaxation
is needed after removing springs because the neutral network has $f_{ij}=0$ in each contact.}

\begin{figure}[t]
\includegraphics[width=8.4cm]{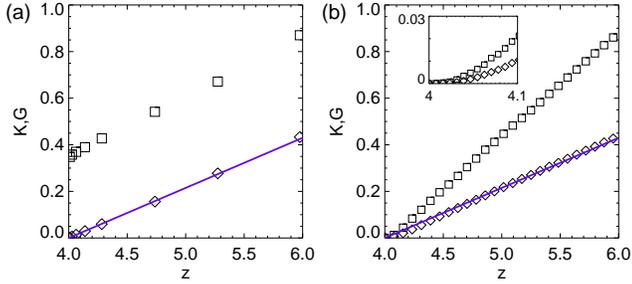}
\caption{Shear (diamonds) and bulk (squares) elastic moduli as function of
distance to jamming for (a) jammed harmonic packings (b) randomly cut
spring networks. Both panels have the same linear fit for the shear modulus (blue line).
{Inset: enlargement of the behavior of $K$ and $G$ near $z=4$.}}
\label{fig:GK}
\end{figure}

\section{Elastic Moduli}
To analyze the linear response, we impose an 
infinitesimal strain deformation of order $\varepsilon$, implemented
by the appropriate changes in rest lengths of all bonds that cross the
boundary of the periodic box {of size $L\times L$.
This amounts to replacing $\mathbf{u}_{ij}$ in 
eq.~(\ref{dynmatdef}) by $\mathbf{u}_{ij}-\mathbf{u}^\mathrm{b}_{ij}$,
where $\mathbf{u}^\mathrm{b}_{ij}$ denotes the appropriate shift of magnitude
$\varepsilon L$ at bonds $ij$ that cross the boundary, and is zero for interior bonds.
Keeping track of this substitution in going from eq.~(\ref{dynmatdef}) to eq.~(\ref{responseeq}), these
constant terms}
are taken to the right hand side, and thus act like an effective
$f^\mathrm{ext}$~\cite{respprl} that is proportional to $\varepsilon$.
The response of the system to
this shape or volume change of the box is then calculated by
solving equation~(\ref{responseeq}) for this effective external
force.

The moduli are extracted from the energy (eq.~(\ref{dynmatdef2})) according
to
\begin{equation}
\label{bulkmod} K,G=\frac{\Delta E}{2V\varepsilon^2}~,
\end{equation}
for a uniform strain, 
$\varepsilon_{xx}=\varepsilon_{yy}=\varepsilon$ for compression, and
$\varepsilon_{xy}=\varepsilon$ for shear.
Here $V$ is the volume of the system.

In fig.~\ref{fig:GK} we show the scaling of the elastic moduli $G$
and $K$ thus obtained, as a function of the contact number $z$ for
both packing-derived and randomly cut spring networks. The main
point is that these two families exhibit different scaling
behavior: for randomly cut networks, both moduli vanish as $\Delta
z$, while for the packing-derived networks only the shear modulus
$G$ goes to zero --- the compression modulus $K$ remains finite\footnote{Here
and in what follows, $\Delta z=z-z_\mathrm{c}\approx z-z_\mathrm{iso}$, where
for packing-derived (randomly cut) networks $z_\mathrm{{c}}=4~(4.045)$. The discrepancy
between $z_\mathrm{c}$ and $z_\mathrm{iso}$ {(see inset of fig.~\ref{fig:GK})} for the randomly cut networks is not a finite size effect, but can
be attributed to the precise cutting protocol.}.
The behavior of the randomly cut networks is consistent with what
is expected for rigidity percolation in random spring networks~\cite{thorpe95,thorpe96},
while the behavior for packing-derived networks is in
agreement with earlier data for jammed packings~\cite{epitome,respprl}. Hence, from the point of view of
rigidity percolation, the anomaly in jammed packings is thus that
the compression modulus $K/k$ stays finite as $\Delta z \to 0$.

Note that the dynamical matrix of both types of networks contains only
geometric information about the network, since the spring constant $k=1$ for
each bond. Hence the crucial difference between packing-derived networks and
randomly cut networks that is causing the bulk modulus to be different must
have a purely geometric origin.

\begin{figure}[t]
\includegraphics[width=8.4cm]{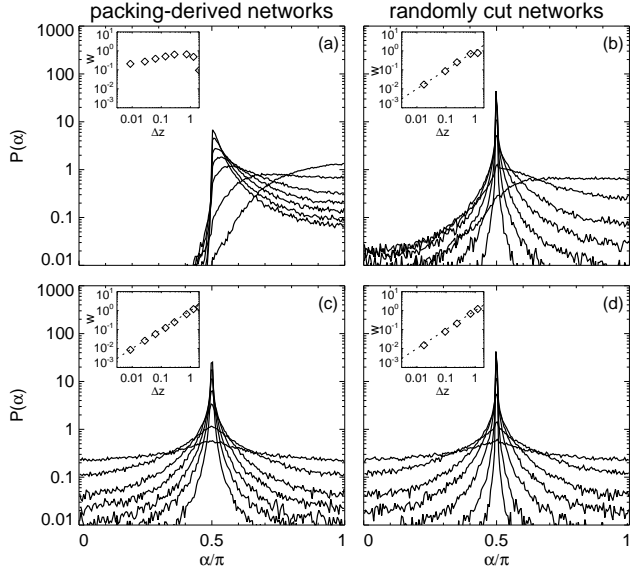}
\caption{The PDF of the displacement angles $P(\alpha)$ for
compression (a,b) and shear (c,d).
The seven curves denote, in order of decreasing
peak height at $\alpha=\pi/2$,
$z=4.008,4.027,4.063,4.14,4.28,4.74,5.27$.  
(a,c) For packing-derived
networks, $P(\alpha)$ for compression and shear appear rather
different. (b,d) For randomly cut networks, $P(\alpha)$ develops the
same peaked structure when $z \rightarrow 4$ for compression and
shear. Insets: The width of the
peaks (defined as the width of the interval containing the central 50\% of the
data), as a function of $\Delta z$.
The dotted lines indicate $w\sim\Delta z$ for all cases except compression of bead packings.}
\label{fig:pa}
\end{figure}

\section{Nonaffinity of Response}
We will now connect the scaling of the elastic moduli to the local
deformation field. One tool that we use to probe the degree of
non-affinity of the response near point $J$  is $P(\alpha)$, the
probability density function (PDF) of the displacement angles
$\alpha_{ij}$~\cite{respprl}. Here $\alpha$ denotes the angle
between $\vc{u}_{ij}$ and $\vc{r}_{ij}$, or,
\begin{equation}
\label{defalpha}
\tan\alpha_{ij}=\frac{u_{\perp,ij}}{u_{\parallel,ij}}~.
\end{equation}

In EMT, the displacements of the particles are prescribed by an
affine deformation field. Affine compression corresponds to a
uniform shrinking of the bond vectors, i.e. $u_{\perp,ij}=0$ while
$u_{\parallel,ij}=-\varepsilon r_{ij}<0$:
the corresponding $P(\alpha)$ exhibits thus
a delta peak at $\alpha=\pi$. The effect of an affine shear on a
bond vector depends on its orientation, and for isotropic random
packings $P(\alpha)$ is flat.

The results for packing-derived networks are shown in fig.~\ref{fig:pa}ac.
Note that far away from jamming, the PDFs are similar to the EMT
predictions: a peak at $\pi$ under compression, and a flat PDF
under shear.
When approaching the unjamming transition, a
peak at $\alpha=\pi/2$ develops, which signifies that an increasingly
large fraction of contacting particles mostly slide past each other. However,
under shear, this peak is much more pronounced than under compression,
and under compression the PDF retains a significant shoulder between $\pi/2$ and $\pi$.

The results from the randomly cut networks are shown in
fig.~\ref{fig:pa}bd:  a strong peak develops in $P(\alpha)$ as
$\Delta z$ decreases, \emph{both for the response to shear and to
compression}. The relative displacements of contacting particles
in response to compression thus signal an important difference
between packing-derived networks and random networks.

\section{Scaling Arguments for Non Affinity}
\begin{figure}[bt]
\includegraphics[width=8.4cm]{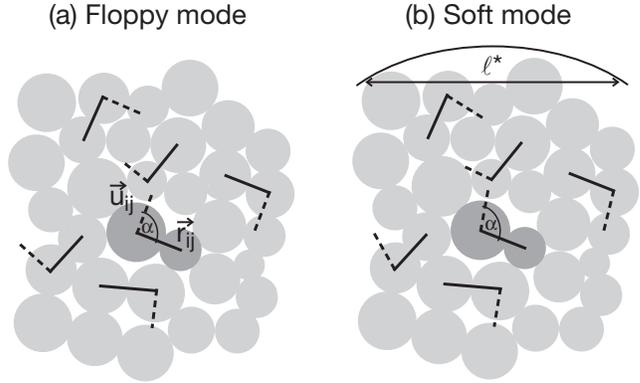}
\caption{Illustration of Wyart's argument~\cite{wyart08} for $\uparl/\uperp$: Left: a floppy
patch of material, obtained by cutting bonds on the outer edge, in which all
contacts have {exactly} $\alpha=\pi/2$ upon distortion{, in accord with
the definition of a floppy mode~\cite{alexander}}. Right: a
{\emph{weakly distorted} floppy mode (also called trial soft mode)
can be thought of as a floppy mode that is distorted elastically on a scale $\ell^*$.
Accordingly all angles $\alpha$ are slightly different from $\pi/2$, as indicated in the figure.}
The dashed lines denote the relative displacement pairs of contacting particles,
marked by the solid line connecting their centers.}
\label{fig:wzz}
\end{figure}
Wyart and coworkers have given arguments for estimating the energies and local deformations
of soft (low energy) modes starting from purely floppy (zero energy) modes~\cite{wyartEPL,wyart08}.
They construct trial soft modes that are basically floppy modes,
obtained by cutting bonds around a patch of size $\ell^*$, and then modulating these with
a sine function of wavelength $\ell^*$ to make the displacements vanish at the 
locations of the cut bonds.
Here $\ell^*\sim 1/\Delta z$ is a characteristic length scale~\cite{wyartEPL, wyartPRE, wyartthesis,respprl}.
In particular, for the local deformations (see fig.~\ref{fig:wzz}), they find~\cite{wyart08}
\begin{equation}
\frac{\uparl}{\uperp}\sim \frac{1}{\ell^*} \rightarrow
\frac{\uparl}{\uperp} \sim \Delta z ~, \label{scale_prediction1}
\end{equation}
where symbols without indices ${ij}$ refer to typical or average
values of the respective quantities%
\footnote{In earlier work~\cite{respprl}, we have argued that the
scaling ${\uparl}/{\uperp} \sim \Delta z$ can also be understood
by balancing the first and second terms in the energy expansion
(eq.~(\ref{dynmatdef})) which yields the scaling
${\uparl}/{\uperp}\sim\sqrt\delta$, with $\delta $ is the typical
overlap. For jammed packings, where the pre-stress term is taken
into account, this result is consistent with
(\ref{scale_prediction1}) in view of the scaling $\Delta z \sim
\sqrt{\delta}$. However, as we show here, even if the
pre-stress term is ignored in the dynamical matrix, very similar
scaling is obtained.}.
Note that the width $w$ of the peak in $P(\alpha)$ is, close to the
jamming transition, roughly $w\sim\uparl/\uperp$, because $|\alpha_{ij}-\pi/2|\approx
u_{\parallel,ij}/u_{\perp,ij}$ if $u_{\parallel,ij}\ll
u_{\perp,ij}$.

The question is now, whether the linear response follows this prediction
for the soft modes, for our two families of networks.  
The insets of fig.~\ref{fig:pa} show that the scaling behavior
(\ref{scale_prediction1}) is consistent with our data for the
width $w$ of the peak of $P(\alpha)$ for packing-derived networks
under shear, and for randomly cut networks under either
compression and shear deformations. The $P(\alpha)$ for
compression of packing-derived networks is the exceptional case.
The peak of $P(\alpha)$ does not grow as much, and a substantial
shoulder for large $\alpha$ remains even close to jamming: the
tendency for particles to move towards each other remains much
more prominent under compression.

\section{Fraction of Compressed Bonds}
In order to clarify the significance of the large-$\alpha$
shoulder, let us analyze the fraction of  significantly compressed
bonds. Intuitively, it is clear that this fraction should be at
the root of the difference between randomly cut networks, whose
compression modulus $K$ vanishes near jamming, and packing-derived
networks whose $K$ does not. Indeed, suppose we compress a
packing-derived network. When a \emph{finite} fraction of the bonds gets
shortened with a finite fraction of the strain $\varepsilon$, then
$K$ will be proportional to the bond strength $k$
--- this simply follows from the expression for the energy change
$\Delta E$, eq.~(\ref{dynmatdef}).

To quantify this, we define the fraction $\rho_\mathrm{comp}$ of
bonds whose local response has $\alpha > 3\pi /4$, i.e.,
contact pairs which upon compression move more towards each other
than they move sideways ($\uparl <-|\uperp | <0$).
If the PDF $P(\alpha)$ was governed by a \emph{single} scale, the
observed scaling of the width of the peak of $P(\alpha) \sim
\Delta z $ in accord with (\ref{scale_prediction1}), suggests that
$\rho_\mathrm{comp}\sim \Delta z$  near jamming.

\begin{figure}[tb]
\includegraphics[width=8.6cm]{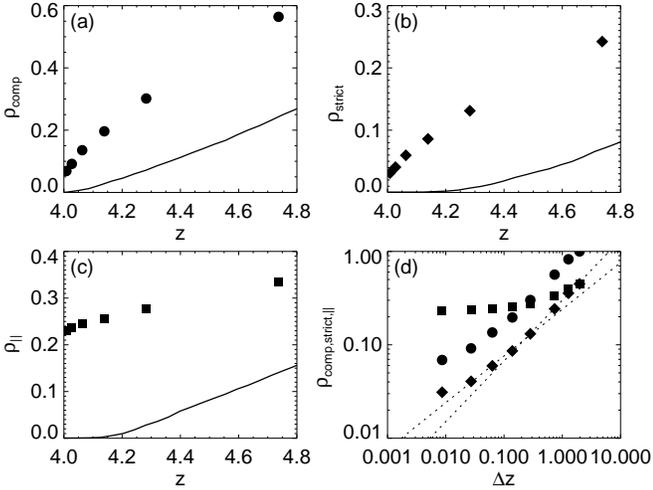}
\caption{Scaling of (a) $\rho_\mathrm{comp}$,
(b) $\rho_\mathrm{strict}$, and (c) $\rho_\parallel$, as a
function of $z$ for compression of packing-derived networks
(symbols) and randomly cut networks (curves). (d) The packing data
on a log scale, emphasizing the rapid rise of $\rho_\mathrm{comp}$ at small $\Delta z$.
The dotted lines mark exponents 0.5 and 0.65, to guide the eye --- there is no clean
scaling.}
\label{fig:rho}
\end{figure}

In fig.~\ref{fig:rho} we compare the scaling of $\rho_\mathrm{comp}$
for both our types of networks.
The $P(\alpha)$ for random networks can be described by a single
scale $1/\ell^*\sim w\sim\Delta z$ (fig.~\ref{fig:pa}b), and indeed the corresponding
$\rho_\mathrm{comp}$ is linear in $\Delta z$ (fig.~\ref{fig:rho}a).
For packing-derived networks close to
jamming, $\rho_\mathrm{comp}$ rises more rapidly than linearly, and
is much larger than for the randomly cut networks. This shows
that under compression of packings a significant fraction of
the contacts remains
non-sliding and that single-parameter scaling does not apply ---
indeed, while our randomly cut networks \emph{are} consistent with a
linear variation of $\rho_\mathrm{comp}$, if we fit our data for
packing-derived networks to a power law form $\rho_\mathrm{comp} \sim (\Delta
z)^\zeta $, we
do not find a clear scaling ($\zeta \approx 0.65$, but only over 1 decade in $\Delta z$).

In principle, many of the bonds with $\alpha > 3 \pi/4$ could have
anomalously small $\uparl$ --- to check that this is not the case,
we have also investigated $\rho_\mathrm{strict}$, the fraction of
bonds whose local response has $\alpha > 3\pi /4$ and $\uparl<
-\varepsilon r_{ij}$, and $\rho_\parallel$, the fraction of bonds whose local
response has $\uparl< -\varepsilon r_{ij}$. The latter condition can be phrased as
the fraction of bonds that are more compressed than they would be if the response were affine.
As shown in fig.~\ref{fig:rho}b-c, these measures of compressed bond fractions 
are also much larger for compression of packing-derived networks.

For compression of packing-derived networks, since $K/k$ remains
finite for $\Delta z \rightarrow 0$,
one should
expect a finite fraction of bonds with $\uparl$ of order
$\varepsilon$ --- consistent with this $\rho_{\parallel}$ remains
finite in this case. Although the $P(\alpha)$ in fig.~\ref{fig:pa}a do not
appear to be governed by a single scale, a tentative argument can be given
why the rise in $\rho_\mathrm{comp}$ is steeper than linear for small $\Delta z$:
Assume the typical $\uperp$ is still of the order $\varepsilon/\sqrt{\Delta z}$, as
is the case for compression of randomly cut networks (from combining eq.~(\ref{scale_prediction1})
with eq.~(\ref{dynmatdef}) and $K\sim k\Delta z$).
Then, the relevant scale in $P(\alpha)$ would be set by $\uparl/\uperp\sim\sqrt{\Delta z}$, and one
would find $\rho_\mathrm{comp}\sim\sqrt{\Delta z}$.
As expected, we do not find such a clear scaling in fig.~\ref{fig:rho}d, but the 
rapid initial rise is clearly visible.

In conclusion, we find that the non-affine displacements in random
spring networks and sheared jammed packings all share the same
simple scalings of $P(\alpha)$, as well as having elastic moduli
which scale as $k \Delta z $, where $k$ denotes the bond
stiffness. The response of jammed packings to compression is the
exceptional case: $P(\alpha)$ has more structure than a single
peak, naive scaling breaks down and $K\sim k$.

\section{Interpretation in terms of the {space of force networks}}
We finally briefly discuss these issues within the framework developed by
Wyart~\cite{wyartthesis,wyart08moduli} for the response of frictionless
granular packings. For a network consisting of $N$ particles and $zN/2$
contacts, any imposed deformation can be expressed in terms of the change of
the rest lengths of  some bonds in the network.  After perturbing one or more
bonds, for example in a way which corresponds to a global shear or compression
of the packing, there will be an energy minimization involving the $dN$ degrees
of freedom (displacements $\vc{u}_i$). Hence, the space of responses to
perturbations that cost energy has dimension $zN/2-dN=\Delta zN/2$. An
equivalent way to view this is that after perturbing the rest lengths of the
bonds, the particles will move so as to satisfy the $dN$ local equations of
force balance.  Therefore the force response network can be expressed in a
basis $\{\vc{f}^{(i)}\}$ of the $\Delta zN/2$-dimensional solution space
$\mathcal{F}$ of of the force balance equations.

The force space thus defined is very similar to the solution space
of the \emph{force network ensemble}~\cite{fneprl,pret,fneshearprl}, where one
studies the space of allowed force configurations,
$\mathcal{F}^\mathrm{fne}$ for a given contact geometry and
externally imposed pressure. Let us define the extended force
network ensemble, as the ensemble of all allowed force
configurations, without the  constraint that the pressure be fixed~\cite{fneshearprl}.
This force space is precisely the $\Delta z N/2$-dimensional space
spanned by the orthonormal basis $\{\vc{f}^{(i)}\}$ defined above.

Now, if we fix the pressure, this leads to an additional
constraint. By a simple rotation in force space it is possible to
choose the $\{\vc{f}^{(i)}\}$ such that $\vc{f}^{(1)}$ precisely gives
the direction of increasing pressure, so that all other base
vectors are perpendicular to the pressure direction~\cite{fneshearprl} --- the force ensemble with fixed pressure
simply results from projecting out the $\vc{f}^{(1)}$ direction
from $\mathcal{F}$.

Suppose we externally impose changes in the rest lengths of the
bonds, denoted by $\vc{y}$ --- for a compression, we may for
example increase all rest lengths.
Wyart~\cite{wyartthesis} then shows that the energy change corresponding
to such external forcing can be expressed as
\begin{equation}
\label{wyartDeltaE} \Delta E=\frac12\sum_{i=1}^{N\Delta
z/2}\langle\vc{f}^{(i)}|\vc{y}\rangle^2~{,}
\end{equation}
{where $\langle\cdot|\cdot\rangle$ denotes the scalar product in force space.}
If we consider a deformation $\vc{y}$ of which the components are of
order $\epsilon$ then in general we may assume that there will be no correlation between $\vc{y}$ and
the force space $\mathcal{F}$. The dominant contribution to the
squared inner product in eq.~(\ref{wyartDeltaE}) is then
$\sum_m(f^{(i)}_m)^2y_m^2\sim \sum_m \epsilon^2/N\sim \epsilon^2$,
and summing these over all basis vectors gives
$ \Delta E\sim N \Delta z \epsilon^2$,
making the energy extensive and proportional to the distance to
the jamming transition times the square of the strain. This is the
case for example if $\vc{y}$ represents a shear deformation, so
that $G\sim\Delta z$.

In this scenario, however, the response to compression is an
exceptional case. Compression amounts to the special situation
that $\vc{y}$ is essentially pointing in the \emph{same} direction
as  the basis vector $\vc{f}^{(1)}$, which we chose to be in the
direction of increasing pressure. In this case, all terms in the
inner product $\sum_m f^{(1)}_my_m$ are positive, and of order
$\epsilon/\sqrt{N}$, so that
$\langle\vc{f}^{(1)}|\vc{y}\rangle^2\sim N\epsilon^2$. Now the
contribution of the first term in eq.~(\ref{wyartDeltaE}) is
already extensive by itself, \emph{and independent of $\Delta z$}.
In this picture, the compression response is anomalous
since it corresponds to an alignment with a special direction in
the force space; in other words, from this point of view too, it
is best to think of compression, and not shear, as being the anomalous response!

\section{Conclusion} 
Since the jamming transition is about loss of rigidity, it is natural to compare
the elasticity of disc packings to rigidity percolation on a random network~\cite{bolton}.
From this reference point, what is special about jammed
packings is that they have an anomalously large resistance to compression.
In addition, the response of packings to compression does not follow the
simple scaling relations that govern the particles' relative displacements
under shear. This interpretation should be contrasted with the more
commonly stated conclusion from comparing to effective medium theory, which
would be that the resistance to shear is anomalously
small~\cite{maksePRL,epitome}.

The anomalous behavior of the bulk modulus has a purely geometric origin, as we
have shown by numerically extracting it from the dynamical matrix, and interpreting
the results in terms of the solution space of the \emph{force network ensemble}, both of
which are purely geometric: It is all a matter of how purely repulsive particles arrange
themselves when they are pressed together.

\begin{acknowledgments}
We thank E. Somfai for providing the numerical code to construct
the granular packings, and M. Wyart and V. Vitelli for enlightening discussions.
WGE and ZZ acknowledge support from the FOM foundation, and MvH
from the Dutch science foundation NWO through a VIDI grant. WGE
thanks the Aspen Center for Physics, where part of this work was
done, for its hospitality.
\end{acknowledgments}


\end{document}